# Quantum coherent negative bend resistance in InSb mesoscopic structures

N. Goel, T. Jayasekera, K. Mullen, M.B. Santos

*Homer L. Dodge Department of Physics and Astronomy,
and Center for Semiconductor Physics in Nanostructures
The University of Oklahoma, 440 West Brooks, Norman OK 73019*

K. Suzuki, S. Miyashita

*NTT Basic Research Laboratories
3-1 Morinosato Wakamiya, Atsugi-shi, Kanagawa 243-0198, Japan*

Y. Hirayama

*NTT Basic Research Laboratories
3-1 Morinosato Wakamiya, Atsugi-shi, Kanagawa 243-0198, Japan
SORST-JST, 4-1-8 Honmachi, Kawaguchi, Saitama 331-0012, Japan*

**Abstract**

Transport measurements were made on four-terminal devices fabricated from InSb/Al$_x$In$_{1-x}$Sb quantum well structures at temperatures from 1.5 to 300K. Negative bend resistance, which is characteristic of ballistic transport, was observed in devices of channel widths 0.2 or 0.5 μm. We have improved upon the existing implementations of R-matrix theory in device physics by introducing boundary conditions that dramatically speed convergence. By comparison with R-matrix calculations, we show that the experimental observations are consistent with quantum coherent transport.

As semiconductor devices shrink in size, they are governed by different physical processes. When the dimensions of a device become less than the mean free path electron transport enters the "classical ballistic" regime where boundary scattering starts to dominate. When device features are smaller than the wavelength of the carriers, electrons are in a "quantum ballistic" regime where coherent dynamics become important. Negative bend resistance (NBR) can be modeled with quantum or classical ballistic transport. Quantized conductance in a quantum point contact [1] requires coherence in the region of the point contact.

Coherent transport may persist up to high temperatures in InSb since this material has the smallest electron effective mass ($m^* = 0.0139\ m_0$) and the highest room-temperature electron mobility of all binary III-V semiconductors. The two-terminal conductance of a narrow constriction (quantum point contact) in a two-dimensional electron gas in InSb at low temperature has been reported earlier [2]. In this paper we report the experimental observation of ballistic transport through NBR in Si-δ-doped InSb/Al$_x$In$_{1-x}$Sb quantum well structures at temperatures up to 185 K. We show that these observations are consistent with quantum coherent transport.

In a four-terminal structure, bend resistance is defined as $R_B = V_{14}/I_{23}$, where $V_{14}$ is the voltage across terminals 1 and 4 and $I_{23}$ is the positive current passed from terminal 2 to terminal 3 (see inset to Fig. 1b). When the mean free path of the electrons is longer than the channel length, electrons injected from terminal 3 will travel in straight lines (i.e. ballistically) to terminal 1. This continues until enough charge accumulates in terminal 1 to deflect the electrons to terminal 4. This produces a potential difference between terminals 1 and 4 ($V_{14}$) that is negative and hence $R_B < 0$ (NBR).

An experiment on GaAs four-terminal devices was earlier modeled using a semi-classical billiard ball model [3]. However due to its smaller $m^*$, electrons in InSb are more likely to follow a quantum mechanical (QM) model. In a QM system, the transport properties are related to the transmission coefficients of electrons through the Landauer-Buttiker formula [4], which gives the bend resistance in a four-terminal device as,

$$R_B = \frac{h}{2e^2}\frac{T_{41}T_{21} - T_{31}^2}{S} \qquad (1)$$

where S is given by [5],

$$S = (T_{21} + T_{41})[(T_{21} + T_{31})^2 + (T_{41} + T_{31})] \qquad (2)$$

Here the $T_{ij}$ are the transmission coefficients of electrons to lead j from lead i. In the semi-classical approach, the $T_{ij}$'s are calculated by a classical billiard ball model. In contrast we calculate $R_B$ assuming the electron transport is ballistic and can be modeled by a single electron Schrodinger equation. The transmission coefficients of the electrons in the system are calculated by R-matrix theory [6] and $R_B$ is then determined by Eqn. (1).

A single InSb quantum-well structure with $Al_{0.09}In_{0.81}Sb$ barrier layers was grown on a semi-insulating GaAs (001) substrate. The sample layer sequence above the substrate is described elsewhere [7]. For characterizing the mobility and density of the quantum-well structures, Hall bar devices with channel widths of 50 μm were defined through photolithography and wet chemical etching. Ohmic contacts were made by thermal evaporation of Indium on contact pads and subsequent annealing at 225 °C for 5 m in a $N_2$ (80%)/ $H_2$ (20%) environment. The longitudinal and the transverse (Hall) resistances as functions of magnetic field at ~1.5 K observed in a sample of van der pauw geometry are given in Fig. 1a. The density (n) and mobility (μ) for electrons in the quantum well at 1.65 K were measured to be 2.4 x $10^{11}$ cm$^{-2}$ and 1.1 x $10^6$ cm$^2$/Vs. At this temperature, we deduce a mobility mean free path of 0.84 μm. To demonstrate ballistic transport, devices with four terminals were defined on several Hall bars by electron beam lithography and reactive ion etching. The etch depth of the trenches that separate the terminals was ~ 0.23 μm. A scanning electron microscope image of a four-terminal structure is shown in the inset to Fig. 1b. Hall resistance through the constriction, given by $R_H$ (= $V_{42}/I_{13}$), at 1.5 K for a lithographic 0.2 μm-wide terminal structure is shown in Fig. 1b.

Fig. 2a shows experimental $R_B$ at 1.5 K as a function of magnetic field for four-terminal structures with a 0.2 or 0.5 μm lithographic device width (*w*). NBR clearly indicates ballistic transport in an InSb quantum well. The NBR amplitude decreases with increasing *w*. The magnetic field bends the trajectory of the electrons away from terminal 1 and the voltage becomes less negative. The 0.2 μm device was pinched off in the second cooling cycle suggestive of nearly 0.1 μm wide depletion widths around the edges of the terminals. A QPC of width 0.2 μm has ~6 occupied states below the Fermi level implying that a QM treatment of a four-terminal device with similar proportions is reasonable [7]. Fig. 2b shows the calculated bend resistance at zero temperature for devices with different widths. The width of the model-device is less than the lithographic widths and the assumed density is 1.9 x $10^{11}$cm$^{-2}$. Both effects arise from depletion, consistent with the results of Fig. 1. At zero temperature, all the energy levels below the Fermi energy are completely full and the ones above are empty, thus $T_{ij} \rightarrow T_{ij}(E_F)$. Since at zero magnetic field, the coefficient $T_{21} > T_{31}$, a large bend resistance is obtained. Our calculation shows that, when the junctions have sharp corners and the subband energy equals or in integral multiples of the subband energies, the difference in $T_{21}$ and $T_{31}$ becomes higher and thus the negative bend resistance is larger. With wedge shaped corners, the sharp resonance features in the transmission coefficients disappear and we see a large bend resistance only for devices with small widths.

The bend resistance of a 0.5 μm- wide four-terminal structure as a function of perpendicular magnetic field at various temperatures is shown in Fig. 3a. As the temperature is raised the NBR amplitude decreases, a consequence of the reduction of the average mean free path. Nonetheless, NBR is observed up to 185 K. We believe the observation of NBR at still higher temperatures is masked by the presence of a lower mean free path, parallel conduction channel in the structure [7]. To calculate $R_B$ at T > 0, we use the effect of Fermi function broadening. The result is shown in Fig. 3b. According to the calculations, the amplitude of $R_B$ becomes less negative as the temperature is increased. Since there are no sharp resonances for the transmission for this geometry, our calculation does not show a strong effect on temperature. The presence of sharp resonances would have increased the contribution to the transport from the electrons near the Fermi energy with the increase in temperature. Furthermore, at some temperature the transport is no longer coherent and a semi-classical treatment is required.

While the theoretical results are comparable to the experimental measurements, uncertainty in the confining potential and the exact carrier density preclude quantitative comparison. The theoretical calculations allows for three qualitative predictions: (1) a device with straight terminal leads rather than wedge shaped ones will show a larger NBR, (2) it is possible for the temperature dependence of the NBR to be non-monotonic if there are sharp resonances in transmission amplitude and (3) if the feature size is made slightly smaller we should see structures arising from the quantum wire subband spacing.

We thank Michael Morrison for his help in applying R-matrix theory to devices. This work was supported by the National Science Foundation under Grant Nos. DMR-0520550 and DMR-0510056.

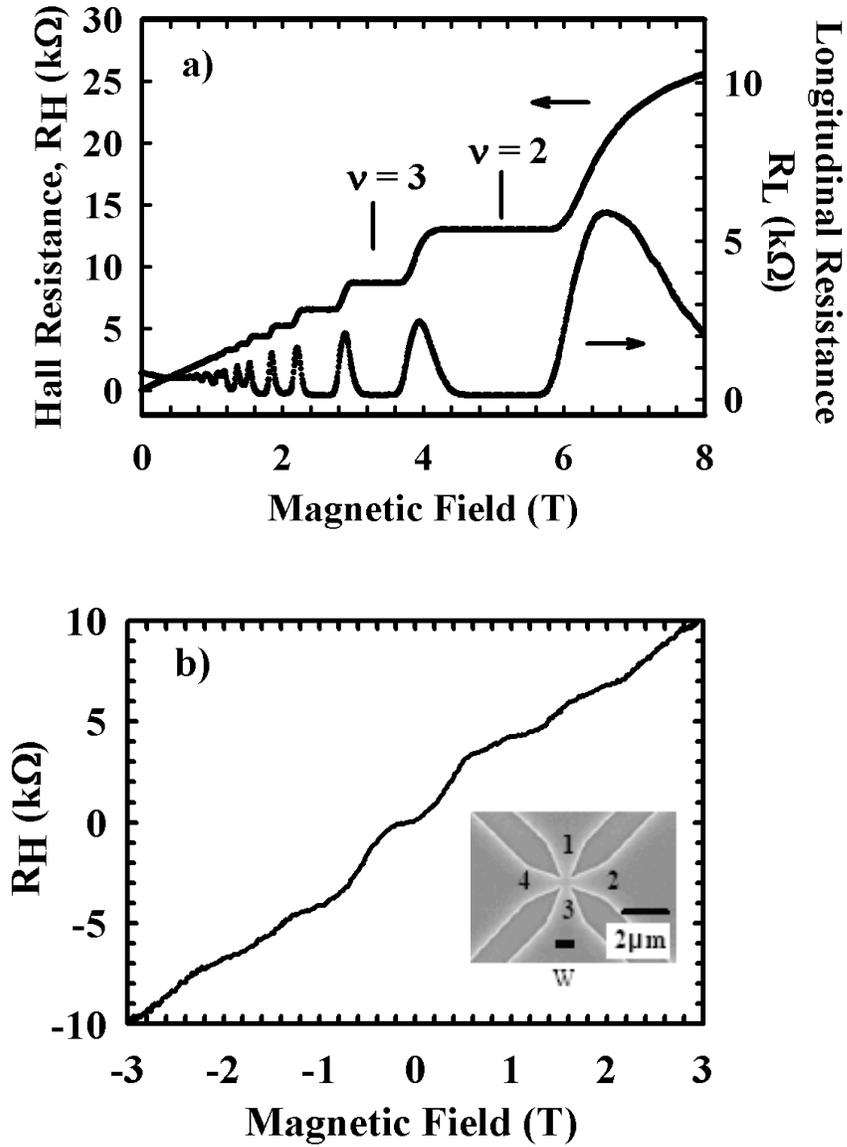

FIG. 1: (a) Hall resistance and longitudinal resistance in InSb single quantum well with $Al_{0.09}In_{0.91}Sb$ barriers as measured on a Hall bar at 1.65 K. b) $R_H (= V_{42}/I_{13})$ as a function of B-field at 1.5 K for a lithographic 0.2 μm-wide four-terminal structure fabricated from an $InSb/Al_{0.09}In_{0.91}Sb$ quantum well.

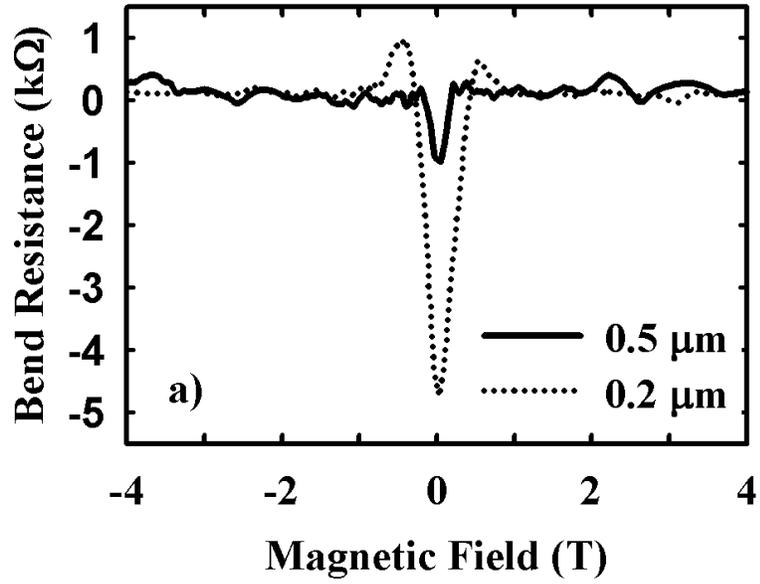

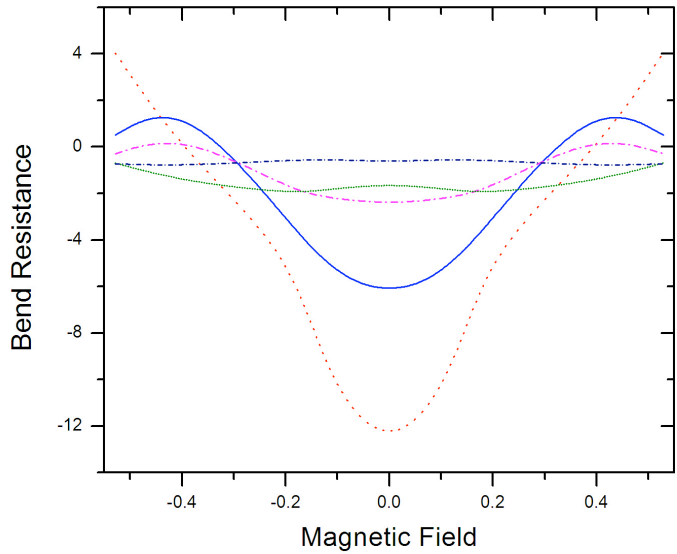

FIG. 2: (a) Experimental bend resistance as a function of magnetic field, B, for two four-terminal structures with 0.2 and 0.5 μm lithographic device width. (b) Calculated bend resistance as a function of B for various device widths (dot- 0.038 μm, solid- 0.045μm, long dash-dot- 0.052 μm, dash- 0.077 μm and short dash-dot- 0.113 μm). The calculation considers the device geometry as shown in the inset to (b) and electron density of $1.9 \times 10^{11}$ cm$^{-2}$.

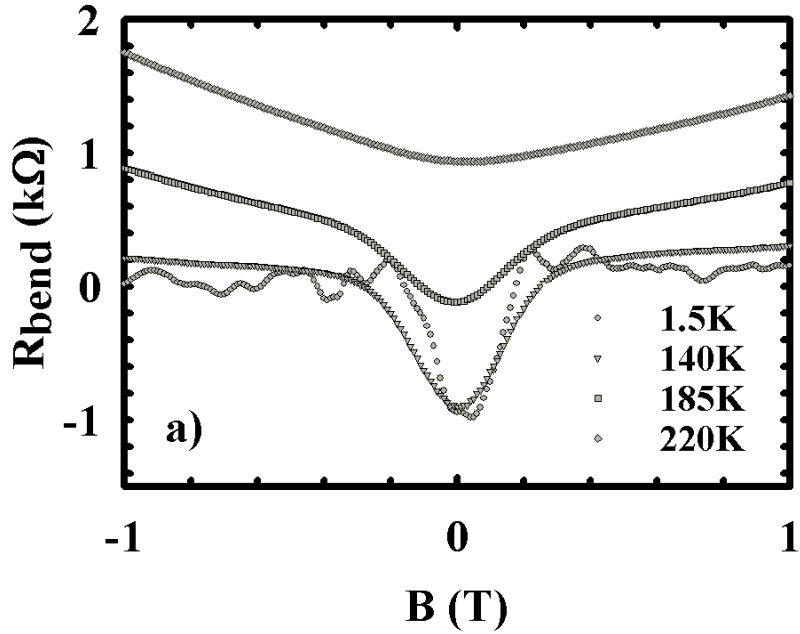

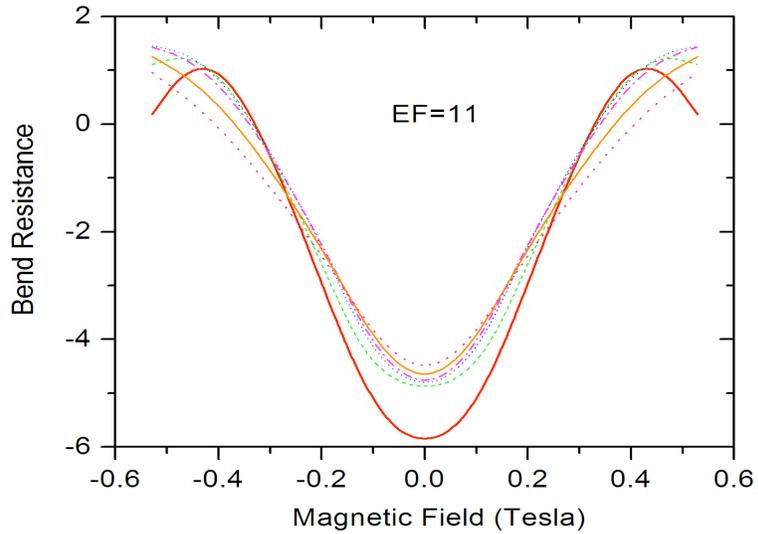

FIG. 3: Experimental (a) and calculated (b) bend resistance as functions of the magnetic field and temperature. The lithographically defined device width for the experimental curve is 0.5 μm. The calculation assumes a device width = 0.045 μm. Different curves are for different temperatures (Line- 4 K, dash- 40 K, Dot- 75 K, Dash-dot- 100 K, solid 150 K, dot 200 K).